\documentclass[useAMS,usenatbib,a4]{mn2e}
\usepackage{amsmath,amssymb}
\usepackage{epstopdf,graphicx,color}
\newcommand{\bc}{}
\title[Characterising Thermal Sweeping]{Characterising thermal sweeping: a rapid disc dispersal mechanism}
\author[Owen et al.]{James E. Owen$^{1}$\thanks{E-mail: jowen@cita.utoronto.ca}, Mathias Hudoba de Badyn$^{1,2}$, Cathie J. Clarke$^{3}$ \& Luke Robins$^{3}$\\
$^{1}$Canadian Institute for Theoretical Astrophysics, 60 St. George Street, Toronto, M5S 3H8, Canada.\\
$^{2}$Department of Physics and Astronomy, University of British Columbia, Vancouver, V6T 1Z1 Canada.\\
$^{3}$Institute of Astronomy, Madingley Rd, Cambridge, CB3 0HA, UK.
}
\begin{document}

\pagerange{\pageref{firstpage}--\pageref{lastpage}} \pubyear{2002}

\maketitle

\label{firstpage}

\begin{abstract}
We consider the properties of protoplanetary discs that are undergoing inside-out clearing by photoevaporation. In particular, we aim to characterise the conditions under which a protoplanetary disc may undergo `thermal sweeping', a rapid ($\lesssim 10^{4}$ years) disc destruction mechanism proposed to occur when a clearing disc reaches sufficiently low surface density at its inner edge and
where the disc is unstable to runaway penetration by the X-rays. We use a large suite of 1D radiation-hydrodynamic simulations to {\bc probe} the observable parameter space, which is unfeasible  in higher dimensions. These models allow us to determine the surface density at which thermal sweeping will take over the disc's evolution and to evaluate this critical surface density as a function of X-ray luminosity, stellar mass and inner hole radius. We find that this critical surface density scales linearly with X-ray luminosity, increases with inner hole radius and decreases with stellar mass and we develop an analytic model
that reproduces these results. 

This surface density criterion is then used to determine the evolutionary state of protoplanetary discs at the point that  they become unstable to destruction by thermal sweeping. We find that transition discs created by photoevaporation will undergo thermal sweeping when their inner holes reach $20-40$ AU, implying 
that  transition discs with large holes {\bc and no accretion} (which were previously
a predicted outcome of the later stages of all flavours of
photoevaporation model) will not form. Thermal sweeping thus avoids
the production of large numbers of large, non-accreting holes (which
are not observed) and implies that the majority of holes created by
photoevaporation should still be accreting. 

We  emphasise that the surface density
criteria that we have developed apply to {\it all} situations 
where the disc develops an inner hole that is optically thin to
X-rays.  It  thus applies not only to the case of holes 
originally created by photoevaporation but also  to holes formed,
for example, by the tidal influence of planets.
\end{abstract}

\section{Introduction}
Understanding the dispersal of protoplanetary discs remains a key problem in star and planet formation. In particular, the time at which the disc disperses sets the time in which (gas) planets must form. Additionally, the method and time-scale on which disc clearing operates strongly influences the dust and gas physics in the disc, as well as the evolution of currently forming planets.

Observationally, optically thick primordial discs are known to survive for several Myr before being dispersed in both dust (Haisch et al. 2001; Hernandez et al. 2007; Mamajek 2009) and gas (Kennedy \& Kenyon 2009), leaving behind a pre-main sequence star possibly surrounded by an optically thin debris disk, containing second generation dust particles (Wyatt 2008). A small fraction of young stars show evidence for a cleared gap or hole in their dust discs, indicated by a lack of opacity at NIR wavelengths, but values comparable with an optically thick disc at MIR wavelengths. These `transition' discs (Strom et al. 1989; Sturski et al. 1990; Calvet et al. 2002) have been interpreted as discs caught at a stage which is intermediate between disc bearing and disc-less. Furthermore, the lack of objects observed in regions of parameter space that represent discs that are no longer optically thick at IR wavelengths indicates that the time-scale for disc dispersal is short compared to the disc's lifetime (Kenyon \& Hartmann 1995). Comparing the ratio of observed transition discs to primordial discs indicates that the clearing time-scale for the inner disc in roughly 10\% of the disc's lifetime (Luhman et al. 2010; Ercolano et al. 2011; Koepferl et al. 2013). Furthermore, the lack of objects which show no NIR or MIR excess but show excesses at FIR or sub-mm/mm wavelengths (Duvert et al. 2000; Andrews \& Williams 2005;2007; Cieza et al. 2013) indicate that this inner disc clearing is correlated with outer disc clearing. 

While `transition' discs have several possible origins such as grain growth (Dullemond \& Dominik 2005; Birnstiel et al. 2012), tidal truncations by planets (Calvet et al. 2005; Rice et al. 2006; Zhu et al. 2012, Clarke \& Owen 2013) or photoevaporation (Clarke et al. 2001), it is planet formation (Armitage \& Hansen 1999) or photoevaporation that offer the best scenarios for disc dispersal. While no qualitative or quantitative predictions can be made about how planet formation clears the disc, one would assume it slows as it proceeds through the disc due to the longer planet formation time and large mass-reservoir to remove.  Photoevaportion driven by X-rays from the central star can explain a large fraction of the observed transition discs (Owen et al. 2011, 2012) but not the full sample; in particular transition discs with large holes and high accretion rates (Espaillat et al. 2010; Andrews et al. 2011) are inconsistent with a photoevaporative origin. However, this category of  `transition' disc may have a different origin and not even be associated with the disc clearing process  at all (Owen \& Clarke 2012; Clarke \& Owen 2013; Nayakshin 2013) being instead perhaps
a relatively rare but  long-lived phenomenon.

Photoevaporation can  explain the population of observed transition discs with small holes ($R_{\rm hole}<20$ AU) and modest accretion rates ($\dot{M}_*<10^{-8}$ M$_\odot$ yr$^{-1}$). However,  the same model predicts a comparable sample of non-accreting transition discs  with large holes; termed `relic discs' by Owen et al. (2011), and a similar population may be expected if planet formation is the dominant dispersal mechanism. While the exact population of transition discs with large holes is yet to be quantified, the lack of Weak T Tauri stars (WTTs) with sub-mm detections (Duvert et al. 2000; Andrews \& Williams 2005;2007, Mathews et al. 2012) or {\it Herschel} detections (Cieza et al. 2013) is inconsistent with a sizeable (comparable to the population of `transition' discs) population of relic discs.

 The 1D viscous calculation of disc dispersal typically employed in disc evolution calculations (Clarke et al. 2001; Alexander et al. 2006; Alexander \& Armitage 2009; Owen et al. 2010,2011) 
neglect the fact that the disc may be unstable to  dynamical clearing, where the disc's mid-plane can be entirely penetrated by the X-rays. Owen et al. (2012) presented simulations of a process they termed  `thermal sweeping' whereby a disc with an optically thin inner
hole is dynamically cleared (on a  time-scale $\sim 10^{3}$ years) once
the surface density at the disc's inner edge falls to a sufficiently low value. 
In the low mass discs simulated by Owen et al, this process set in at  radii $\sim 10-20$ AU.
 Owen et al. (2012) argued that this process prevents the production of a long-lived
population of low mass discs with large cavities (`relic discs') and that it
shifts the predicted populations of transition discs created through photoevaporation to those dominated by small holes and low accretion rates. 

In addition, Owen et al. (2012) presented a back of the envelope calculation to estimate the surface density $\Sigma_{TS}$ at which thermal sweeping begins; however, such a calculation was only compared to two multi-dimension radiation hydrodynamic calculations. A parameter span using multi-dimensional calculations is not currently computationally feasible. Thus, in this paper, we build on the framework for thermal sweeping outlined in Owen et al. (2012) in order to determine the critical surface density for thermal sweeping to proceed by performing a large set of 1D radiation-hydrodynamic calculations. In Section~2 we describe the thermal sweeping mechanism and summarise the back of the envelope calculation performed by Owen et al. (2012). Section~3 outlines the numerical method used for suite of 1D radiation hydrodynamic calculations and present our findings in Section~4. In Section~5 we discuss the implications for thermal sweeping on disc evolution and discuss an improved model for determining the critical density at which thermal sweeping proceeds and finally in Section~6 we summarise our findings. 
 
\section{Basics of thermal sweeping}
In this section we describe the theoretical basis for how X-ray irradiation of a disc with an inner hole leads to rapid,   dynamical disc dispersal. We also present the back of the envelope calculation presented by Owen et al. (2012) as well as pointing out the assumptions that may break down.
\subsection{Understanding thermal sweeping}
\begin{figure}
\centering
\includegraphics[width=\columnwidth]{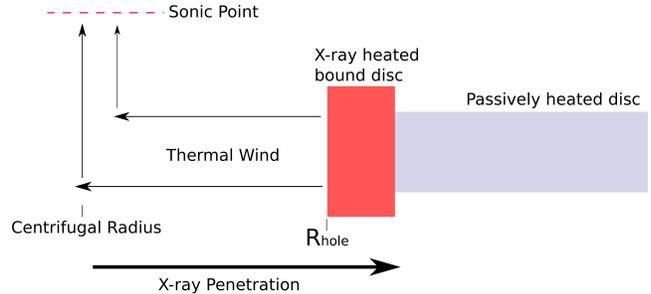}
\caption{Schematic diagram of the flow/disc structure resulting from X-ray irradiation of a disc with a cleared inner hole.}\label{fig:cartoon}
\end{figure}

In order to explain the origin of thermal sweeping, we
show in Figure~\ref{fig:cartoon} a  schematic depiction
of the flow structure that results from X-ray irradiation of a disc with an inner hole, based on the simulations of Owen et al. (2010,2011,2012).

The X-rays heat up the inner disc and drive a sub-sonic wind radially inwards. 
The simulations of Owen et al. (2012) showed that the gas becomes
significantly decelerated by centrifugal effects
\footnote{Recall that viscous effects are negligible in the dynamical
wind flow and thus the gas in this region conserves its specific
angular momentum} 
at
 a radius of roughly $R_{\rm hole}/2$
and at this point is deflected vertically
by pressure gradients, attaining a sonic transition at a height
$z \sim R$.

  Owen et al (2012) showed that the mass flux in the wind is set at the
sonic point and that the structure of the radially inflowing
thermal wind region (see Figure 1) can be understood in terms
of a one-dimensional, angular momentum conserving flow subject
to this boundary condition on the mass flux. The total radial column density
in this thermal wind region is given by:

\begin{eqnarray}
N&\sim& nR_{\rm hole}= \frac{\dot{M}_w}{2\pi\mu v R (H/R)}\nonumber \\
&=&10^{21}{\rm \, cm}^{-2}\left(\frac{\dot{M}_w}{10^{-8}{\rm \,M}_\odot{\rm \, yr}^{-1}}\right)\left(\frac{v}{10^{6}{\rm \, cm \,s}^{-1}}\right)^{-1}\nonumber \\
&&\times\left(\frac{R}{10 {\rm \, AU}}\right)^{-1}\left(\frac{H/R}{0.1}\right)^{-1}
\end{eqnarray}
where $N$ is the column, $n$ is the number density, $\dot{M}_w$ the mass-loss rate, $\mu$ the mean particle mass, $v$ the gas velocity and $H$ the scale height.
This is 
significantly lower than the column required to absorb the X-rays ($\sim10^{22}$ cm$^{-2}$ - Ercolano et al. 2008;2009). 
Thus, in this flow topology the material that dominates the absorption of the X-rays is not actually in the thermal wind, but is still bound to the disc, essentially in hydrostatic equilibrium with temperatures $<1000$ K. 

Owen et al. (2012) argued that for this bound X-ray heated region to remain in dynamical equilibrium it must have a radial pressure scale ($\Delta\equiv{\rm d}R/{\rm d}\log P$) smaller than its vertical pressure scale (${\rm d}z/{\rm d}\log P\equiv H=c_s/\Omega$). {\bc Since in order to obtain dynamical equilibrium, the disc must be able to adjust itself radially to changes in the vertical structure on a time-scale considerably shorter than the dynamical time-scale of the disc $(\sim 1/\Omega)$}.  In situations where $\Delta\sim H$
(and where the time-scales for attaining
equilibrium in the radial and vertical directions are comparable), 
this bound X-ray heated region is unstable to progressive penetration by the X-rays as follows. In the case 
of a small vertical expansion of the flow,
the disc cannot re-gain equilibrium in the radial direction on the
vertical expansion time-scale and so the radial column density at
the mid-plane (and the local volume density) are reduced. Consequently,
the temperature of the X-ray heated flow  rises and the vertical expansion
is accelerated. The further reduction in mid-plane column density
results in further penetration of the X-rays and the process of {\it
vertical} evacuation of the disc is repeated for a fresh layer of
previously unheated disc material. The key difference between
this behaviour and `normal' photoevaporative clearing from
a holed disc is that vertical expansion leads to runaway heating
of the disc inner rim.

\subsection{Column limited estimate for thermal sweeping}
Here we repeat the simple analytical argument presented by Owen et al. (2012), but emphasise the assumptions made in its derivation. Since the X-ray bound heated region must be in approximate pressure equilibrium with the passively heated (by dust) disc at large radii one can write:
\begin{equation}
k_bn_XT_X=P^{\rm dust}_h\label{eqn:disc_pressure}
\end{equation}
where $k_b$ is Boltzmann's constant, $n_X$ \& $T_X$ is the number density and temperature in the bound X-ray heated region and $P_h^{\rm dust}$ is the gas pressure in the passively heated disc. Writing $\Delta$ as approximately $N_X/n_X$, and using the fact that thermal sweeping begins when $\Delta=H$, we can recast Equation~\ref{eqn:disc_pressure} in terms as a criterion for thermal sweeping as:
\begin{equation}
P_h^{\rm dust}\le\frac{N_Xk_bT_X}{H}
\end{equation}
where $N_X$ is the radial column density corresponding to
complete absorption of the X-rays ($\sim 10^{22}$ cm$^{-2}$). 
In the case of a disc in vertical hydrostatic equilibrium this can
be re-cast in 
terms of a critical surface density for the inner rim of the dust-heated disc:
\begin{equation}
\Sigma_{TS}= 0.4 {\,\rm g\,\,  cm}^{-2}\left(\frac{T_X}{400\,K}\right)^{1/2}\left(\frac{T_{\rm dust}}{20\,K}\right)^{-1/2}\label{eqn:o12_crit}
\end{equation}
Given $T_{\rm dust}\propto R^{-1/2}$, this resulted in a weak $R^{1/4}$ scaling for the critical surface density with radius, with  no explicit dependence on stellar mass or X-ray luminosity. 

 However, this derivation presented by Owen et al. (2012) makes two strong assumptions: i) that the bound X-ray heated region always represents a column density of $10^{22}$ cm$^{-2}$; ii) the surface density at the transition between the X-ray heated disc and passively heated disc is representative of the peak surface density of the disc's profile. 
 
Assumption i) may break down if the gas density is so high that X-ray heating is insufficient to heat the gas above the dust temperature even if it is optically thin to the X-rays. This typically occurs at ionization parameters $\xi_{\rm min}\lesssim 10^{-7}$ erg cm s$^{-1}$ (where $\xi=L_X/nr^2$). In the case that the boundary of the X-ray heated region
is set by the criterion $\xi=\xi_{\rm min}$, 
we can estimate the column density in the bound X-ray heated region as $N=n\Delta$. At the point of onset of thermal sweeping this can be
written $N=(H/R)nR$ which (from the definition of the ionization parameter) becomes:
\begin{eqnarray}
N&=&10^{22} {\rm\, cm}^{-2}\left(\frac{H/R}{0.1}\right)\left(\frac{L_X}{10^{30} {\rm\, erg\, s}^{-1}}\right)\nonumber\\&&\times\left(\frac{R}{10{\rm \, AU}}\right)^{-1}\left(\frac{\xi_{\rm min}}{10^{-7}{\rm \, erg\,cm\,s}^{-1}}\right)^{-1}
\end{eqnarray}
Therefore, since we expect thermal sweeping to take place when the hole radius is large ($R_{\rm hole}\gtrsim 10$ AU), then we need to consider
 cases where the total column density in the bound X-ray heated region is $<10^{22}$ cm$^{-2}$. This means that 
Equation~\ref{eqn:o12_crit} will not represent how the critical surface density for thermal sweeping varies in certain regions of the parameter space. In particular, if the criterion in some regions of parameter space is set by a minimum ionization parameter rather than total column density then it should depend on the X-ray luminosity as well as inner hole radius.

 In this paper we assume without further discussion that
$\Delta=H$ is a necessary and sufficient condition for initiating
thermal sweeping (an assumption that we will revisit in future
two-dimensional studies motivated by the results of the present
paper.) 
   Here  we use 1D radiation-hydrodynamic models to perform a parameter space study,  calculating how the critical surface density ($\Sigma_{TS}$ - defined as the peak surface density at the inner edge of the dust-heated disc
for which  $\Delta=H$) varies with stellar mass, X-ray luminosity and inner hole radius.
The advantage
of a one-dimensional treatment is that it is possible to
perform a large suite of high resolution simulations that are
beyond the scope of 2D simulations and that can inform
the choice of parameters and resolution requirements of future
2D studies. The justification for a one dimensional treatment
is that - prior to the onset of thermal sweeping - the
flow is to a good approximation one dimensional
in the vicinity of the inner edge.

\section{Numerical Method}
In order to calculate the inner disc structures we must make use of numerical hydrodynamic calculations. We use the {\sc zeus} code (Stone et al. 1992; Hayes et al. 2006), which has been modified by Owen et al. (2010;2012) to include the heating of circumstellar material by stellar X-ray photons. We use this modified code to calculate the mid-plane inner disc structure; therefore, we run the code radially in one-dimension ({\bc although we also evolve the azimuthal velocity}\footnote{Often referred to as 1.5D simulations.}). In the absence of X-ray heating the disc's temperature is set to the mid-plane temperature expected for a passively irradiated protoplanetary disc (e.g. Chiang \& Goldreich 1997; D'alessio et al. 2001):
\begin{equation}
T_{\rm mid}=\max\left[T_{\rm 1 AU}\left(\frac{R}{1{\rm \,\,AU}}\right)^{-1/2}, 10 {\rm \,\,K}\right]
\label{eqn:cg}
\end{equation}
where $T_{\rm 1 AU}$ is the mid-plane temperature at 1 AU, and we adopt typical values of 100 K for a 0.7 M$_\odot$ star and 50K for a 0.1 M$_\odot$ star (D'alessio et al. 2001). As initial conditions we adopt a $\Sigma\propto R^{-1}$ profile, where in order to convert between surface density and mid-plane gas density we assume the disc is vertically isothermal and use:
\begin{equation}
\Sigma(R)=\sqrt{2\pi}H(R)\rho(R)
\end{equation} 
This mid-plane density profile is setup in hydrostatic equilibrium including the necessary modification to the Keplerian rotation due to the radial pressure gradient. 
\begin{figure}
\centering
\includegraphics[width=\columnwidth]{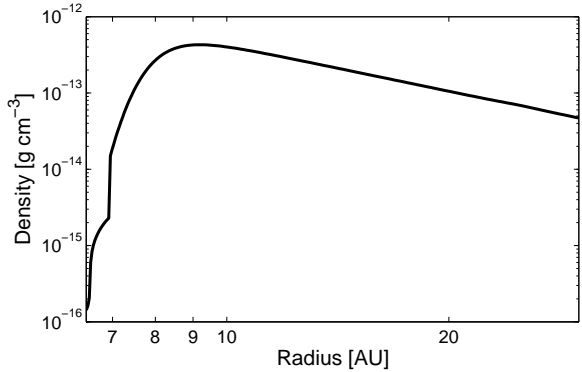}
\caption{Density structure for a simulation of a disc with an inner hole around a 0.1 M$_\odot$ star with an X-ray luminosity of $2\times10^{30}$ erg s$^{-1}$. We note the outer boundary is not shown and is at a radius of 38 AU in this calculation.}\label{fig:den_flow}
\end{figure}
\begin{figure}
\centering
\includegraphics[width=\columnwidth]{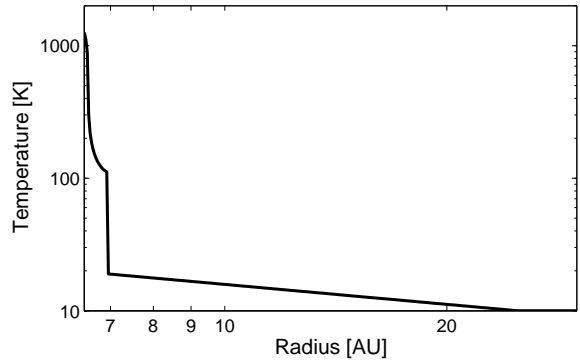}
\caption{Temperature structure for the same simulation shown in Figure~\ref{fig:den_flow}.}\label{fig:temp_flow}
\end{figure}

In order for the simulations to obtain stable configuration the boundary conditions require careful consideration. The outer boundary condition is implemented in order to maintain hydrostatic equilibrium across the outer boundary throughout the simulation, so sub-sonic radial flows are not allowed to interact with the boundary. Explicitly we do this by maintaining a $\Sigma\propto R^{-1}$ power law across the boundary and adjust the azimuthal velocity in the boundary cells in order to maintain this equilibrium.  In X-ray heated winds, the mass-flux is determined explicitly by the conditions at the sonic surface (Owen et al. 2012; Owen \& Jackson 2012).  It is not possible to
follow the flow to the sonic point
in a one dimensional simulation since this occurs at a
height $Z \sim R$ above the disc. Therefore
  we adjust the inner boundary condition in order to match the flow properties (density, temperature, velocity) onto the results of an on-axis streamline taking from the multi-dimensional inner hole simulations performed by Owen et al. (2012). Such a method is sufficient provided the inner boundary lies between the centrifugal radius and inner hole radius (so the problem remains remains essentially 1D, see Figure~\ref{fig:cartoon}). 
 
 The radial grid is  non-uniform, spanning $[R_i,6R_i]$ - where $R_i$ in the radius of the inner boundary - with a high resolution of 200 cells; typically 
this gives $\sim 10-15$ cells per radial pressure scale height in the inner hole region. In general, the structures are spatially resolved with a lower resolution, typically 100 cells; however, the higher resolution reduces the scatter in properties calculated (particularly $\Delta$). Under this set-up the inner disc is then irradiated, with the X-rays driving a flow off the inner disc with a mass-flux specified by the conditions at the inner boundary; this flow slowly erodes material from the disc resulting in an inner hole radius that increases slowly with simulation time. The evolution is slow enough that the structure maintains a quasi-dynamical equilibrium at each time-step. We stop the simulation once the inner hole moves to such a radius that the inner boundary is no longer outside the centrifugal radius, and the 1D approximation is no-longer appropriate. In Figure~\ref{fig:den_flow} \& \ref{fig:temp_flow} we show an example of the quasi-steady state flow structure obtained from one of the simulations, for a disc around a 0.1 M$_\odot$ star with an X-ray luminosity of $2\times10^{30}$ erg s$^{-1}$.

Using 1D simulations allows us to probe {\bc a large range of the} observable parameter space of disc models at high resolution. We perform roughly $5000$ simulations covering a range of initial surface densities and  radial ranges for two stellar masses. For the 0.1 M$_\odot$ star we consider X-ray luminosities\footnote{\bc Although low-mass stars may have lower X-ray luminosities, the simulated range is sufficient to calibrate the model discussed in Section~5, which can be applied to the full range of X-ray luminosities.} of $2\times 10^{29}$ and $2\times10^{30}$ erg s$^{-1}$ and hole sizes in the range  $\sim 1-20$. For the 0.7 M$_\odot$ star we consider X-ray luminosities of $2\times10^{29}$, $2\times 10^{30}$ and $2\times10^{31}$ erg s$^{-1}$ with inner hole sizes in the range $\sim 10-70$ AU. For each simulated radial range we vary the initial surface density normalisation at intervals of $0.15$dex over the range expected in clearing discs. These simulations are then used to compute how the ratio of $\Delta/H$ varies with surface density, thus extracting the critical surface density for thermal sweeping and how it varies varies with stellar mass, X-ray luminosity and inner hole radius.

\section{Results}
\begin{figure}
\centering
\includegraphics[width=\columnwidth]{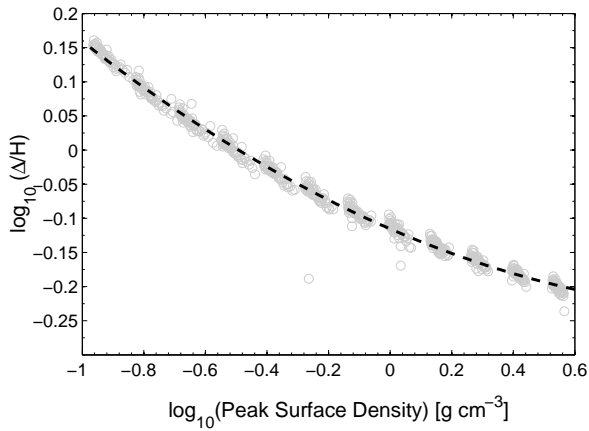}
\caption{The variation in the ratio of $\Delta/H$ with peak surface density for a disc with an inner hole at $\sim 10$ AU around a 0.7 M$_\odot$ star and X-ray luminosity of $2\times10^{30}$ erg s$^{-1}$. The points represent individual values computed at a given time in the simulation, where the scatter is indicative of the resolution. The dashed line shows a polynomial fit to the points. Points which sit far from the line occur at early time-steps where the disc has not reached quasi-dynamical equilbrium.}\label{fig:ratio_surf}
\end{figure}

Our suite of simulations cover a large range of parameter space
and provide   the mid-plane structure of discs with large inner holes. 
For a given inner hole radius and
surface density normalisation in the disc we use the simulation to
solve for the ratio
of $\Delta$ to $H$ and can hence determine how this ratio varies
as a function of the maximum surface density in the dust heated disc.
An example of how $\Delta/H$ varies with the peak surface density in a disc is shown in Figure~\ref{fig:ratio_surf}.

Figure~\ref{fig:ratio_surf} demonstrates how we use the simulation data to calculate $\Sigma_{TS}$. We use the simulations to regularly store the current flow state (density, temperature, velocity); thus for each data dump we compute the ratio of $\Delta/H$, the current hole radius and the peak surface density of the disc. Then for each radial range we string together all these values in order to produce a plot of how this ratio varies with the peak surface density in the disc. These data points necessarily have some scatter due to the finite grid resolution (around 0.05dex for our resolution), and therefore we fit a polynomial to this data series (typically a quadratic). We then use the fitted polynomial to compute the peak surface density in the disc at which $\Delta/H=1$ and the inner hole radius at which this occurs. Then we can combine the simulations at different radial ranges to see how $\Sigma_{TS}$ varies with inner hole radius, X-ray luminosity and stellar mass. 
\subsection{Variation with inner hole radius \& stellar mass}
\begin{figure}
\centering
\includegraphics[width=\columnwidth]{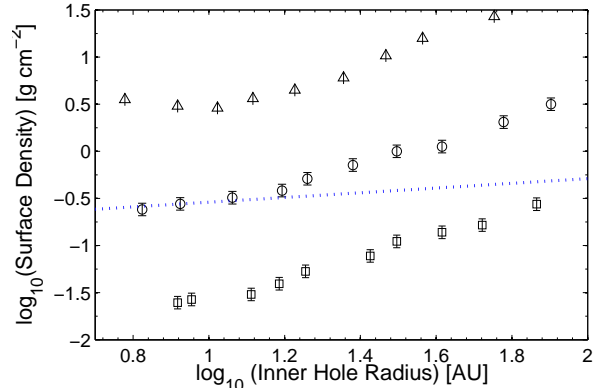}
\caption{Critical surface density for thermal sweeping ($\Delta=H$) shown as a function of inner hole radius around a 0.7 M$_\odot$ mass star for three X-ray luminosities: $2\times10^{31}$ erg s$^{-1}$ (triangles), $2\times10^{30}$ erg s$^{-1}$ (circles)  and $2\times10^{29}$ erg s$^{-1}$ (squares). The error bars show the uncertainity in the calculation of the critical surface density introduced by the finite grid resolution (see text and Figure~\ref{fig:ratio_surf}).{\bc We also show a $\Sigma\propto R^{1/4}$ fit (perfomed at 10~AU) as the dotted line, to represent the column-limited model (Equation~4), which clearly fails to reproduce the simulations.}}\label{fig:crit0.7}
\end{figure}

\begin{figure}
\centering
\includegraphics[width=\columnwidth]{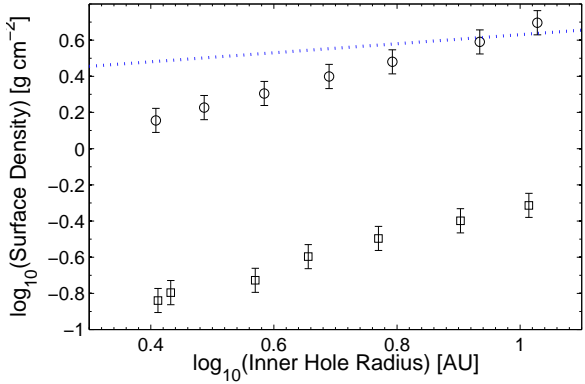}
\caption{Same as Figure~\ref{fig:crit0.7} but shown for a 0.1 M$_\odot$ mass star}\label{fig:crit0.1}
\end{figure}

The resulting variation in the critical surface density for thermal sweeping is shown as a function of inner hole radius and X-ray luminosity in Figure~\ref{fig:crit0.7} \& \ref{fig:crit0.1} for discs around 0.7 \& 0.1 M$_\odot$ solar mass stars respectively. The finite grid resolution that results in the scatter in Figure~\ref{fig:ratio_surf} gives rise to an uncertainty in the surface density of $\sim\pm 0.05$ dex which we represent as error bars in the Figures. We emphasise this scatter does not mean the simulations are not spatially resolved, in-fact the simulations are well resolved (we perform several resolution tests and note the simulations are resolved with $\sim$100 cells, we use 200 cells) and the resolution has been chosen to minimise this uncertainty while allowing a computationally feasible parameter study.

The general trends of the critical surface density with inner hole radius and X-ray luminosity are characterised in Figure~\ref{fig:crit0.7}. In general we find the surface density increases with radius as well as X-ray luminosity. 
  At the highest X-ray luminosity of $2\times10^{31}$ erg s$^{-1}$, we find the surface density is roughly constant at small radius ($R_{\rm hole}<10$ AU). This range represents the column limited case discussed  by Owen et al. (2012)
(equation \ref{eqn:o12_crit}), {\bc where the size of the X-ray heated region is purely set by the absorption of the X-rays and $\Delta$ is simply given by $\sim N_X/n_X$}. However, as we will discuss further in Section~5, over most of parameter space  the 
extent of the X-ray heated region is  set 
by a minimum ionization parameter, rather than an optical depth cut-off.

\section{Discussion and Observational Implications}
In the previous section we have shown how the critical surface density for thermal sweeping varies with stellar mass, inner hole radius and X-ray luminosity. As expected we find that thermal sweeping will proceed once the remaining mass in the disc is low $\lesssim 1$ M$_{\rm Jup}$. However, we do not
  find the shallow $R^{1/4}$ dependence expected from the Owen et al. (2012) model, but a steeper variation with inner hole radius. Furthermore, we find an explicit, roughly linear, dependence on the X-ray luminosity, not predicted by Owen et al. (2012). In Section~2 we outlined the two assumptions that underpinned the previous calculation: that the total column density in the bound X-ray heated layer is $\sim10^{22}$ cm$^{-2}$, and that the surface density at the inner edge of the passively heated disc may not be representative of the peak surface density in the disc. We find that while the second clause certainly plays a role, it is not the major reason why the simulations do not show the expected dependence. It is the first reason: that the total column density in the bound X-ray heated layer is not always $10^{22}$ cm$^{-2}$ and in the majority of realistic cases is lower than this value and a function of radius. 

\begin{figure}
\centering
\includegraphics[width=\columnwidth]{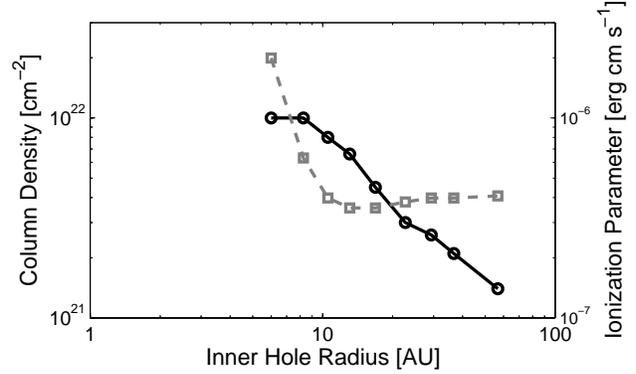}
\caption{The column density (solid line) and ionization parameter (dashed line) in the X-ray heated bound region, shown as a function of inner hole radius for a disc around a 0.7 M$_\odot$ star irradiated at an X-ray luminosity of $2\times10^{30}$ erg s$^{-1}$. The points show the radius of individual simulations.}\label{fig:column_radius}
\end{figure}

For example we show in Figure~\ref{fig:column_radius} how the total column density and the ionization parameter {\bc at the transition from X-ray bound to dust heated layer} varies with inner hole radius for the case of a disc around a 0.7 M$_\odot$ star irradiated at a high X-ray luminosity of $2\times10^{31}$ erg s$^{-1}$. This Figure clearly shows
that, at about $10$ AU the X-ray heated region makes a transition
from being column limited to being ionization parameter limited, or more correctly X-ray temperature limited.


\subsection{An improved model for thermal sweeping}
We can now use the above result: that the properties of the warm X-ray heated bound portion of the disc are set by a minimum ionization parameter rather than total column condition. 
Since the gas velocity in the bound portion of the disc (both dust and X-ray heated) is extremely sub-sonic, we can treat these regions of the disc as being approximately hydrostatic. Inside a radius $R_c$ (i.e. radius
in the disc at which the source of material for the wind is in
centrifugal balance) the specific angular momentum is constant and the effective gravity in the radial direction is given by:
\begin{eqnarray}
g_{\rm eff}&=&-\frac{GM_*}{R^2}\left(1-\frac{R_c}{R}\right)\nonumber\\
&=&\Omega^2\delta(R)\label{eqn:geff}
\end{eqnarray}
where $\delta(R) = R_c - R$ and $\Omega$ is the local Keplerian velocity. Now using the definition that $\Delta$ is the radial pressure scale length in the X-ray heated gas we see that:
\begin{equation}
\Delta\equiv c_X^2/g_{\rm eff}=H_X^2/\delta\label{eqn:Delta1}
\end{equation}
where $c_{X}$ is the sound speed at the outer edge of the X-ray heated
disc and we have applied the condition for hydrostatic equilibrium normal
to the disc plane, i.e. $c_{X} = H_X \Omega$. So on the point of thermal sweeping ($\Delta=H_X$), Equation~\ref{eqn:Delta1} tells us that 
$\delta=H_X$ so that $\delta\ll R_c$. Thus we proceed, in the limit $\delta \ll R_c$ and retain terms in $g_{\rm eff}$
to first order in $\delta/R$:  we solve for radial hydrostatic equilibrium in the dust-heated disc, where the approximation $\delta\ll R_c$ ensures that the dust-heated disc
is approximately isothermal at a temperature $T_c$. (We note that while
$T_c$ varies with radius on a scale length $\sim R_c$, making the isothermal approximation is formally equivalent to only retaining terms first order in $\delta/R$.) This then yields
a Gaussian profile for the dust-heated disc such that

\begin{equation}
P(\delta) = P_c\, \exp\left[ -\left(\frac{1}{2}\frac{\delta}{H_c}\right)^2\right]
\label{eqn:pprofile}
\end{equation}
where $P_c$ is the pressure at $R_c$ and $H_c$ is the hydrostatic
scale height of the disc in the vertical direction at $R=R_c$. 
The innermost extent of the dust-heated disc
is thus set by the condition of pressure balance with the X-ray
heated region  (pressure $P_X$) - which we set as $R_{\rm hole}$ for convenience - and is given by:
\begin{equation}
\delta(R_{\rm hole}) = 2H_c \sqrt {\log\left(\frac{P_c}{P_X}\right)^2}
\label{eqn:delx}
\end{equation}
Applying the condition for thermal sweeping ($\Delta=H_X$) allows Equation~\ref{eqn:pprofile} to be re-written as:
\begin{eqnarray}
\frac{P_c}{P_X}&=&\exp\left[\frac{1}{2}\left(\frac{H_X}{H_c}\right)^2\right]\nonumber\\
&=&\exp\left(\frac{T_X}{2T_c}\right)
\end{eqnarray}

 We can qualitatively understand this result by considering
a case where $T_X/T_c$ is fixed. If $P_c/P_X$ is high then
the X-ray heated gas can only achieve pressure equilibrium
with the cold gas at a point that is relatively far from
$R_c$, so the effective gravity is large (Equation~\ref{eqn:geff}) and
the radial scale length correspondingly small (Equation~\ref{eqn:Delta1}). However,
as $P_c/P_X$ is reduced, the X-ray heated region is pushed closer to $R_c$, where $g_{\rm eff}$ is low and
the radial scale length becomes larger. When $P_c/P_X$ is reduced
sufficiently, the thermal sweeping criterion ($\Delta = H_X$)
is activated. 

\begin{figure}
\centering
\includegraphics[width=\columnwidth]{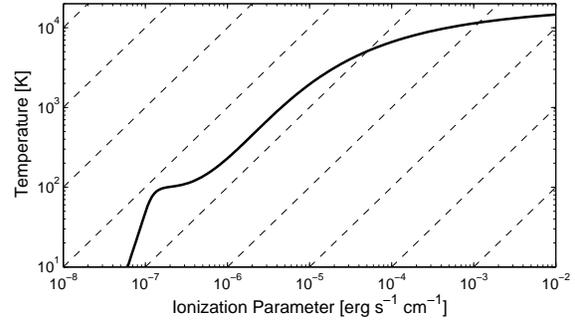}
\caption{Temperature plotted as a function of X-ray ionization parameter, the thin dashed lines show isobars.}\label{fig:txi}
\end{figure}

 This minimum value of $P_c$ can be readily converted
into a minimum surface density criterion since in hydrostatic
equilibrium $P_c \propto \Sigma c_c \Omega$, where $c_c$ is the sound speed at $R_c$. Thus we have:

\begin{equation}
\Sigma_{TS} = \sqrt{2\pi}\frac{P_X}{c_c \Omega} \exp\left(\frac{T_X}{2T_c}\right)\label{eqn:sigts}
\end{equation}
However, Equation~\ref{eqn:sigts} only allows us to evaluate
$\Sigma_{TS}$ (for a given radius in the disc and corresponding
temperature of the dust heated disc) if one also knows 
$T_X$ and $P_X$. It is now necessary to consider the form
of the relationship between $\xi$ and $T$ which we use
to assign X-ray temperatures. In the dense conditions
at the base of the X-ray heated flow (i.e. at low ionisation
parameter), the relationship is nearly flat at a temperature of
$\sim 100$K and then  steepens at $\xi = \xi_{\rm min}$, with
pressure then declining mildly as the temperature is reduced
below $\sim 100$ K. Where the structure of the temperature-ionization parameter relation is determined by the falling X-ray heating efficiency with ionization fraction (Xu and Mccray 1991), in combination with cooling by lines (Owen et al. 2010).  The number density on the cold side of the interface is such that
the X-ray temperature at this point is equal to $T_c$ since
in this case the X-rays would be unable to heat this region
above $T_c$. This condition sets the pressure at the interface;
on the X-ray heated side of the interface the gas shares the same
pressure but has a temperature of $\sim 100$K, since this places
it in the nearly isothermal portion of the $T-\xi$ relationship.
We can see from Figure \ref{fig:txi} that a line of constant pressure
($T \propto \xi$) that originates on the $\xi-T$  curve at
a typical dust temperature of $10-100$ K crosses the curve at a value
of $\sim \xi_{\rm min}$, as indicated by our simulations (see Figure~\ref{fig:column_radius}). We can therefore simply set $P_X$
and $T_X$ by requiring $\xi = \xi_{\rm min}$ at this point, where comparing with Figure~\ref{fig:txi} and our simulations indicate a sensible choice for $\xi_{\rm min}=3\times10^{-7}$ erg s$^{-1}$ cm$^{-1}$. 
Substituting this condition into \ref{eqn:sigts} we obtain:
\begin{eqnarray}
\Sigma_{TS}\!\!\!&=\!\!\!&\sqrt{2\pi}\mu c_c\left(\frac{L_XT_X}{\Omega \xi_{\rm min}R_{\rm hole}^2T_c}\right)
\exp\left(\frac{T_X}{2T_c}\right)\label{eqn:final_sol2}\\
& =&  0.14 \,{\rm g}\, {\rm cm}^{-2}\left(\frac{L_X}{10^{30}{\rm \, erg\, s}^{-1}}\right)\left(\frac{T_{1\rm{AU}}}{100{\rm \, K}}\right)^{-1/2}\nonumber\\
&\times&\!\!\!\left(\frac{M_*}{0.7\,{\rm M}_\odot}\right)^{-1/2}\left(\frac{R_{\rm hole}}{10{\rm \, AU}}\right)^{-1/4}\nonumber\\
&\times &\!\!\! \exp\left[\left(\frac{R_{\rm hole}}{10{\rm\, AU}}\right)^{1/2}\left(\frac{T_{1\rm{AU}}}{100{\rm \, K}}\right)^{-1/2}\right]\label{eqn:final_sol}
\end{eqnarray}
Note that in constructing \ref{eqn:final_sol} from \ref{eqn:sigts} we
have also assumed that the temperature in the
dust-heated disc follows a $R^{-0.5}$ profile (cf equation \ref{eqn:cg}) and $T_{1{\rm AU}}$ is the dust temperature at 1\,AU defined in Equation~\ref{eqn:cg}.

\begin{figure}
\centering
\includegraphics[width=\columnwidth]{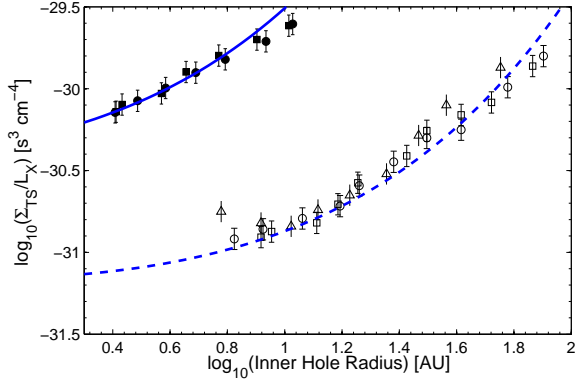}
\caption{Comparision between the thermal sweeping model presented in Equation~\ref{eqn:final_sol}: plotted as $\Sigma_{TS}/L_X$ verus radius, and the simulations presented in Section~4. The open points and dashed line are for the 0.7 M$_\odot$ star and the filled points and solid line are for the 0.1 M$_\odot$ star. The square, circular and triangular points are for X-ray luminosities of $2\times10^{29}$, $2\times10^{30}$ and $2\times10^{31}$ erg s$^{-1}$ respectively.}\label{fig:compare}
\end{figure}

We can compare Equation~\ref{eqn:final_sol} to our simulations by plotting $\Sigma_{TS}/L_X$ for the two sets of simulations around a 0.1 \& 0.7 M$_\odot$ mass star. Given the result for $\Sigma_{TS}/L_X$ presented above has no further dependence on X-ray luminosity, but is only a function of mass and radius then the simulation points should line up along the same line. This comparison is shown in Figure~\ref{fig:compare}, where we plot the simulations (0.7 M$_\odot$ - open points; 0.1 M$_\odot$ - filled points) for the various X-ray luminosities to Equation \ref{eqn:final_sol} (0.7 M$_\odot$ - dashed line; 0.1 M$_\odot$ - solid line). We see the agreement is exceptionally good, for the full range of parameters explored (apart from the small radius runs at the highest X-ray luminosity that are column limited, and are not covered by the model as discussed above).
Figure~\ref{fig:compare} also clearly shows that thermal sweeping is more efficient around lower-mass stars, possibly indicating that it may become ineffective around higher mass stars, as the X-ray luminosity fails to increase with stellar mass around intermediate mass stars. 

Furthermore,  the general form (Equation~\ref{eqn:final_sol2}) could be applied to cases where an optically thin (to the X-rays) gap is created by a planet (e.g. Rice et al. 2006, Zhu et al. 2012) or by a combination of photoevaporation and a planet (Rosotti et al. 2013), to see if planet formation could trigger rapid disc dispersal through thermal sweeping.

\subsection{Properties of observed transition discs}
Confident that our model presented above can accurately predict the surface density of discs that are unstable to rapid clearing by thermal sweeping, we can now make observational predictions for the inner hole radius at which thermal sweeping clears the disc. We use Equation~\ref{eqn:final_sol} to re-analyse the disc population synthesis model performed by Owen et al. (2011), which was designed to match the evolution of disc fraction as a function of time. The model well reproduced a large fraction of transition discs with small holes and low accretion rates, and  fully explains the population of transition discs with low mm fluxes that Owen \& Clarke (2012) identified as discs most likely to be transitioning from disc bearing to a disc-less state. However, in doing so the model also predicted a large population of 'relic' discs with large holes and no accretion. This population is yet to be observed and Owen et al. (2012)
noted that this problem would be alleviated by the thermal sweeping mechanism.

\begin{figure}
\centering
\includegraphics[width=\columnwidth]{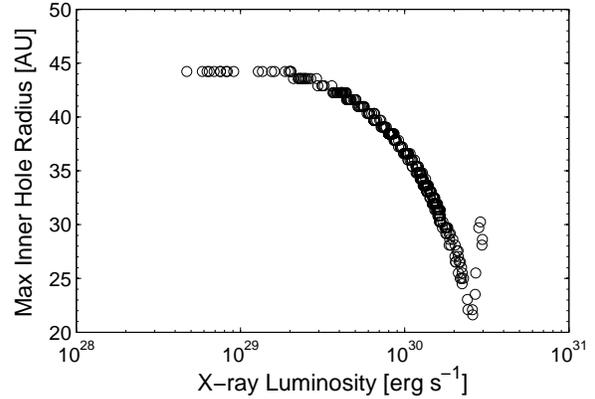}
\caption{Maximum inner hole radii reached by clearing discs before thermal sweeping takes over. Each point represent an individual  disc model taken from the population synthesis calculation of Owen et al. (2011), that followed the evolution of a group of discs evolving under photoevaporation around a 0.7 M$_\odot$ star with an X-ray luminosity drawn from the observed X-ray luminosity function (G\"udel et al. 2007).}\label{fig:max_rad}
\end{figure}

Thus we take the synthetic disc population run by Owen et al. (2011), and assume the disc is dispersed instantaneously by thermal sweeping once the disc's surface density drops below $\Sigma_{TS}$. We determine the distribution of maximum cavity radius for transition discs created by photoevaporation. The resulting radius distribution is shown in Figure~\ref{fig:max_rad}, where we plot the maximum inner hole radius reached before thermal sweeping takes over for discs evolving around a 0.7 M$_\odot$ star with different X-ray luminosities. 

Figure~\ref{fig:max_rad} clearly shows that thermal sweeping prevents the formation of relic discs (i.e those with large holes and no accretion), explaining the lack of a significant population of transition discs observed to have large holes with no accretion. The trend of smaller hole radii reached around higher X-ray luminosity stars is due to the ability of thermal sweeping to begin at higher surface densities with higher X-ray luminosities. At low X-ray luminosities ($< 10^{29}$ erg s$^{-1}$)
the hole radius is {\it independent} of X-ray luminosity: this is
because both $\Sigma_{TS}$ and also the surface density of the disc
at a given radius at the point of photoevaporative hole opening
scale linearly with
X-ray luminosity and so equality is always achieved at
a fixed radius. At larger X-ray luminosities, the process of
`photoevaporation starved accretion' (Drake et al. 2009; Owen et al. 2011) causes a slightly sub-linear
dependence on X-ray luminosity for the disc surface density at hole opening.
Consequently the thermal sweeping radius declines somewhat with
X-ray luminosity. However, at the highest X-ray luminosities this is countered by the fact that gap formation happens at earlier times and hence higher surface densities for stars with higher X-ray luminosities. 

Furthermore, the rapid dispersal of transition discs created by photoevaporation at radii $<45 AU$ means that the ratio between the time spent in an accreting transition disc phase to a non-accreting transition disc phase is greatly enhanced, compared to the photoevaporation model analysed in Owen et al. 2011. For example, our new thermal sweeping models predicts that the median model ($L_X=1.1\times10^{30}$ erg s$^{-1}$ - Figure 9 of Owen et al. 2011), initially opens the hole  at $\sim 2.5$ AU and the disc is destroyed by thermal sweeping when $R_{\rm hole}\sim 30$ AU, with a remaining disc mass of a few Jupiter masses. Furthermore, it spends approximately $2\times 10^{5}$ years as an accreting transition disc and $6.5\times 10^{4}$ years as a non-accreting transition disc. Thus, the addition of thermal sweeping suggests that transition discs created by photoevaporation spend the {\it majority} of their life as accreting transition discs. Due to the variation of thermal sweeping with stellar mass, thermal sweeping can begin earlier around lower mass stars, and may become inefficient around intermediate mass stars, where the X-ray luminosity no longer increases with stellar mass above 1-2M$_\odot$ (e.g. Guedel et al. 2007, Albacete Colombo 2007)

\subsection{Possible limits for the time-scale of thermal sweeping}
One possible limit on the time-scale at which thermal sweeping proceeds through the disc is energetic. The multi-dimensional simulations that were initially used to investigate thermal sweeping enforced local radiative equilibrium. This condition meant energetic consideration were neglected (see discussion in Owen
et al 2010 who demonstrated that for standard X-ray photoevaporation the
mechanical luminosity of the wind is a comfortably small fraction
($\lesssim 8 \%$) of the X-ray luminosity of the source). So $P{\rm d}V$ work done escaping from the potential was not accounted for. The dynamical time-scale on which thermal sweeping begins is not going to be limited by this consideration (because the mass at the inner edge is small). However, as most of the disc material is at large radius, it may take longer than the dynamical time-scale at the inner edge for the disc material at such large radius to be fully heated up to the equilibrium temperature. We can estimate an `energy-limited' time-scale for clearing to occur by comparing the gravitational potential energy stored in the disc, to the rate at which the X-rays inject energy into the system.

We can calculate the gravitational binding energy of a power-law disc of the form  $\Sigma=\Sigma_{\rm hole}(R/R_{\rm hole})^{-1}$ as:
\begin{eqnarray}
U_{\rm disc}&=&\!\!\int_{R_{\rm hole}}^{R_{\rm out}}\!\!\!2\pi R{\rm d}R\frac{GM_*}{R}\Sigma_{\rm hole}\left(\frac{R}{R_{\rm hole}}\right)^{-1}\nonumber\\
&=&\!\!2\pi GM_*\Sigma_{\rm hole}R_{\rm hole}\log\left(\frac{R_{\rm out}}{R_{\rm hole}}\right)\nonumber\\
&\sim &\!\!10^{40}{\rm \, ergs}\left(\frac{M_*}{0.7{\rm \,M_\odot}}\right)\left(\frac{\Sigma_{\rm hole}}{0.3{\rm \, g\, cm}^{-2}}\right)\left(\frac{R_{\rm hole}}{10{\rm \, AU}}\right)
\end{eqnarray}
Comparing this binding energy to the received X-ray luminosity, we find an energy-limited clearing time-scale of:
\begin{eqnarray}
\tau_{\rm clear}&=&\frac{U_{\rm disc}}{\epsilon L_X}\nonumber\\
&=&2\times10^{3} {\rm\, years}\left(\frac{L_X}{10^{30} {\rm \, erg\,s}^{-1}}\right)^{-1}\left(\frac{\epsilon}{0.25}\right)^{-1}\nonumber\\&&\times\left(\frac{R_{\rm hole}}{10{\rm \, AU}}\right) \left(\frac{M_*}{0.7{\rm \,M_\odot}}\right)\left(\frac{\Sigma_{\rm hole}}{0.14{\rm \, g\, cm}^{-2}}\right)
\end{eqnarray} 
where $\epsilon$ represents the fraction of X-rays intercepted by the disc (the X-ray photosphere to the star occurs at height $> H$, Owen et al. 2010). This time-scale is still much shorter than the clearing time-scale for the inner disc by standard photoevaporation $\sim 3\times10^{5}$ years (Clarke et al. 2001, Alexander 2006; Owen et al. 2011). Thus, we expect thermal sweeping to still be efficient in removing the outer disc. Moreover, since $\Sigma_{TS}\propto L_X$, this energy limited time-scale is independent of X-ray luminosity.  However, future multi-dimensional calculations that relax the assumption of radiative equilibrium will be required to fully assess the actual efficiency of the thermal sweeping process at large radius. 

\section{Summary}

In this work we have investigated the rapid disc dispersal mechanism for holed disc `thermal sweeping', introduced by Owen et al. (2012) which takes over from photoevaporation and destroys the remaining disc material on a short ($\lesssim 10^{4}$ year) time-scale once the surface density of the disc has dropped to sufficiently low values. We use the criterion suggested by Owen et al (2012) (involving equality
of the disc's  vertical scale height and the radial thickness of the
X-ray heated bound gas at the rim of the holed disc) in order to define the
point at which a disc will undergo rapid dispersal, using a large suite of 1D radiation-hydrodynamical
simulations. In this way we identify the surface density at which thermal sweeping proceeds as a function of X-ray luminosity, stellar mass and inner hole radius. Our results however do not replicate the analytic estimate presented
by Owen et al 2012 since this was based on the assumption that the
extent of the X-ray heated region at the disc's inner rim is set by
a fixed absorption column. Instead we find (apart from at the
largest X-ray luminosities and smallest radii) that the limits of X-ray
heating are set by the ionisation parameter falling to  the value
$\xi_{min}$, below which the temperature-ionisation parameter relation
steepens strongly (see Figure 8), at which point the X-rays do not
heat the gas above the local disc temperature.  We use this result to derive a new criterion (Equation 15) for how the critical surface density for thermal sweeping varies with physical properties of disc systems and find that
this provides an excellent match to the simulation results (see Figure 9). 

Our main findings are summarised below:
\begin{enumerate}
\item Thermal sweeping can rapidly clear the disc, once the disc mass drops below a few Jupiter masses and the inner hole is at sufficiently large radii $\gtrsim 10$ AU.

\item The critical surface density for thermal sweeping to proceed scales linearly with X-ray luminosity, and increases with inner hole radius.

\item Thermal sweeping is considerably more efficient at clearing discs around lower mass stars and could possibly become ineffective around intermediate mass stars which have comparatively weaker X-ray emission.

\item Using the derived condition for thermal sweeping, we show that transition discs created through photoevaporation are completely destroyed once their inner hole radii reach sizes $>20-40$ AU.

\item The destruction of transition discs by thermal sweeping, prevents the formation of `relic' discs meaning that the {\it majority} of transition discs created by photoevaporation should be found to be accreting, albeit modestly ($\dot{M}_*<10^{-8}$ M$_\odot$ yr$^{-1}$).
\end{enumerate} 

Future multi-dimensional radiation-hydrodynamic simulations are required to understand the details of thermal sweeping. In particular, energy considerations raise questions about how rapid thermal sweeping can be in heating the outer parts of the disc once runaway penetration is in progress. Additionally, penetration and heating of the disc by the FUV radiation field (e.g. Gorti \& Hollenbach 2009; Gorti et al. 2009) may aid in thermal sweeping. The inclusion of FUV heating in hydrodynamic simulations still remains challenging, and we cannot speculate on the role played by FUV heating in thermal sweeping at this stage. 

\section*{Acknowledgements}
We are grateful to the referee for comments that improved the paper. We thank Giovanni Rosotti, Barbara Ercolano, Geoff Vasil, Emmanuel Jacquet and Tom Haworth for insightful discussions. JEO is grateful to hospitality from the IoA, Cambridge during the initial stages of the work. MHB acknowledges support of an NSERC summer research grant held at CITA during 2012. The calculations were performed
on the Sunnyvale cluster at CITA which is funded by the
Canada Foundation for Innovation. We would like to
acknowledge the Nordita program on Photo-Evaporation in Astrophysical
Systems (June 2013) where part of the work for this paper was carried
out.

\end{document}